\documentclass[prd,onecolumn,notitlepage,eqsecnum,nofootinbib,floatfix,aps]{revtex4-1}
\usepackage{amssymb}
\usepackage{amsmath}
\usepackage{amsfonts} 
\usepackage{bm}
\usepackage{graphicx}
\usepackage{comment} 
\usepackage{color}
\usepackage{slashed}
\DeclareSymbolFont{EulerScript}{U}{eus}{m}{n}
\SetSymbolFont{EulerScript}{bold}{U}{eus}{b}{n}
\DeclareSymbolFontAlphabet\scrpt{EulerScript}
\newcommand{\gotht}{\mathfrak{t}}
\newcommand{\goths}{\mathfrak{s}}
\newcommand{\gothr}{\mathfrak{r}}
 
\newcommand{\gothp}{\mathfrak{p}} 
\newcommand{\gothj}{\mathfrak{j}} 
\newcommand{\gothg}{\mathfrak{g}}
\newcommand{\gothf}{\mathfrak{f}}
\newcommand{\gothe}{\mathfrak{e}}
 
\renewcommand{\d}{\mathrm{d}}
\newcommand{\del}{\partial}
\allowdisplaybreaks

\begin{document}
\title{Coulombic contribution to angular momentum flux in general relativity} 

\author{B\'eatrice Bonga}
\email{bbonga@perimeterinstitute.ca}
\affiliation{Perimeter Institute for Theoretical Physics, Waterloo, Ontario, N2L 2Y5, Canada}

\author{Eric Poisson}  
\email{epoisson@uoguelph.ca}
\affiliation{Department of Physics, University of Guelph, Guelph,
	Ontario, N1G 2W1, Canada} 

\date{\today} 

\begin{abstract}
The flux of angular momentum in electromagnetism cannot be expressed entirely in terms of the field's radiative degrees of freedom. Its expression also involves Coulombic pieces of the field, in the form of a charge aspect $q(\theta,\phi)$, a function of polar angles whose integral gives the total charge of the system. Guided by the strong analogy between radiative processes in electromagnetism and gravitation, we ask whether the flux of angular momentum in general relativity might also involve Coulombic pieces of the gravitational field. Further, we ask whether such terms might have been missed in the past by specializing the flux to sources of gravitational waves that are at rest with respect to the frame in which the flux is evaluated. To answer these questions we bring together the Landau-Lifshitz formulation of the Einstein field equations, which provides specific definitions for angular momentum and its associated flux, and the Bondi formalism, which provides a systematic expansion of the metric of an asymptotically flat spacetime in inverse powers of the distance away from the matter distribution. We obtain a new expression for the flux of angular momentum, which is not restricted to sources of gravitational waves at rest nor to periodic sources. We show that our new expression is equivalent to the standard formula used in the literature when these restrictions are put in place. We find that contrary to expectations based on the analogy between electromagnetism and gravitation, the flux of angular momentum in general relativity can be expressed entirely in terms of the field's radiative degrees of freedom. In contrast to electromagnetism, no Coulombic information is required to calculate the flux of angular momentum in general relativity.    
\end{abstract} 

\maketitle
\section{Introduction} 
\label{sec:intro} 

A bounded distribution of electric charges undergoing a dynamical process produces electromagnetic radiation that carries off some of the distribution's energy, linear momentum, and angular momentum. It is natural to expect that the fluxes of energy, linear momentum, and angular momentum should be expressible entirely in terms of the radiative degrees of freedom of the electromagnetic field. In a given gauge, and far away from the source, these can be encoded in a transverse vector potential $A^a_{\rm t}$, where $a$ is a spatial vector index and the label ``t'' indicates that the vector is geometrically transverse, that is, orthogonal to the direction of wave propagation. While this expectation is indeed verified for the fluxes of energy and linear momentum, it is actually false in the case of angular momentum. A recent investigation by Ashtekar and Bonga \cite{ab1,ab2} (see also \cite{bgp}) reveals that in addition to the radiative degrees of freedom, the flux of angular momentum also involves ``Coulombic pieces'' of the electromagnetic field. These are encoded in a ``charge aspect'' $q(\theta,\phi)$, a function of the polar angles $\theta$ and $\phi$ whose integral is equal to the system's total charge. This perhaps unexpected feature of the flux of angular momentum is illustrated in a vivid way by a simple system consisting of a charged sphere that rotates on an axis with a variable angular velocity \cite{bpy}. In this situation the flux of angular momentum is proportional to both the sphere's total charge and the second time derivative of its magnetic moment; it reflects an interplay between radiative and Coulombic pieces of the electromagnetic field.   

Gravitation is strongly analogous to electromagnetism. In the case of gravity, it is well understood that a distribution of masses undergoing a dynamical process produces gravitational radiation that carries off energy, linear momentum, and angular momentum. The analogy suggests that the flux of angular momentum in general relativity might also involve more than just the radiative degrees of freedom of the gravitational field. Is it possible that a dependence on ``Coulombic pieces'' of the field has hitherto been missed?  Given the crucial importance of balance laws (for energy, linear momentum, and angular momentum) in the dynamical modeling of gravitational-wave sources (see \cite{blanchet:14} for example), it appeared to us imperative to find a definitive answer to this question. 

Before revealing this answer, let us flesh out a plausibility argument in favor of a dependence on Coulombic pieces of the gravitational field. First, we argue that Coulombic information may enter the flux of angular momentum on the basis of the strong analogy between gravitation and electromagnetism. Because the electromagnetic flux involves a charge aspect $q(\theta,\phi)$, we might expect that the gravitational flux should involve some kind of mass aspect $m(\theta,\phi)$. A plausible candidate for this is $M(u=-\infty,\theta,\phi)$, the Bondi mass aspect evaluated in the remote past, in the limit in which retarded time $u$ approaches minus infinity. For example, the mass aspect of a boosted source of gravitational waves is given by
\begin{equation}
	M(u=-\infty,\theta,\phi) =  \frac{m}{\gamma^3(1-v\cos\theta)^3},
        \label{eq:massaspect} 
\end{equation}
where $m$ is the source's mass, $v$ is the boost velocity, and $\gamma = (1-v^2)^{-1/2}$. (For simplicity we have taken the boost to be directed along the polar axis, so that the mass aspect does not depend on $\phi$.) This expression was first displayed in  
\cite{BvBM} --- see their Eq.~(72) --- and we provide a derivation in Appendix~\ref{app:boosted}. 

Second, we examine a number of past derivations of the angular-momentum flux in general relativity, and observe that given the assumptions made in those derivations, it is possible that Coulombic information could be missing from the flux. Specifically, derivations based on the Landau-Lifschitz formalism are implicitly or explicitly restricted to sources of gravitational waves that are at rest with respect to the frame in which the flux is evaluated. The expression for the flux that is almost universally used in the gravitational-wave literature is the one displayed in Sec.~IV D of Thorne's seminal {\it Multipole expansions of gravitational radiation} \cite{thorne}, and this expression is based on the the Landau-Lifschitz formalism. In his article, Thorne provides no derivation, but refers to unpublished lecture notes by Bryce DeWitt (which have since been published in \cite{dewitt}); a derivation of the flux that reproduces Thorne's result can also be found in Sec.~12.2.4 of \cite{pw}. In the discussion that follows his statement of the flux, Thorne explains that the formula is meant to apply only to sources at rest;  and this restriction is implicit in the derivation detailed in \cite{pw}. Because the mass aspect $M(u=-\infty,\theta,\phi)$ reduces to the constant $m$ when $v = 0$, it is possible that terms that otherwise would be present in the flux are omitted when the source is at rest. This would leave us with an incomplete description of angular-momentum flux in general relativity. The standard formula might well apply to sources at rest, but it would not be valid in more general situations. In particular, in contexts involving a boosted source, or a source recoiling because of the emission of gravitational waves, the flux would miss terms that incorporate Coulombic information about the gravitational field. 

Another statement of the angular-momentum flux, by Ashtekar and Streubel \cite{AshtekarStreubel}, is  independent of the Landau-Lifshitz formalism, and is not restricted to sources at rest. Their derivation, however, is based on the phase space of radiative modes at null infinity, and it therefore excludes Coulombic information from the start; this expression also could be incomplete. Yet another derivation of the flux of angular momentum, based on the covariant phase-space methods of Wald and Zoupas \cite{WaldZoupas}, was provided by Flanagan and Nichols \cite{FlanaganNichols}. While these methods do not seem to be restricted to the radiative phase space of asymptotically flat spacetimes, and therefore seem to account for Coulombic information, their flux is a time-integrated version of the instantaneous flux considered in this paper, and it is difficult to determine whether the instantaneous flux might depend on the Coulombic terms.  

To find out if anything is indeed missing from the standard formula for the flux of angular momentum in general relativity, we present a new and independent calculation of the flux, taking care to incorporate no assumption regarding the state of motion of the source of gravitational waves. This derivation combines two essential ingredients. The first is the Landau-Lifshitz formulation of the Einstein field equations \cite{landau-lifshitz:b2}, as reviewed in Sec.~6.1 of \cite{pw}. This formulation provides a {\it definition} for the angular-momentum tensor $J^{ab}$ of an asymptotically flat spacetime, and a {\it definition} for its associated flux ${\scrpt T}^{ab}$. Conservation of angular momentum is embodied in the balance law     
\begin{equation}
	\frac{d}{du} J^{ab} = 	-{\scrpt T}^{ab},
	\label{eq:balance}
\end{equation}
which follows as a direct consequence of the field equations. These {\it definitions} are by no means unique, but they are convenient, widely used in the literature, and they provide a firm basis for a discussion of angular-momentum flux. The second ingredient is a systematic expansion of the metric of an asymptotically flat spacetime in inverse powers of $r$, the spatial distance from the matter distribution. This is provided by the Bondi metric \cite{BvBM, Sachs}, which is presented in geometrical coordinates $(u,r,\theta,\phi)$ tied to the expanding light cones of the spacetime.    

Our derivation brings together the Landau-Lifshitz and Bondi frameworks into a happy and fruitful marriage. In this formulation we obtain explicit expressions for $J^{ab}$ and ${\scrpt T}^{ab}$, and verify explicitly that Eq.~(\ref{eq:balance}) follows as a consequence of the field equations. Our expression for the flux reduces to Thorne's standard formula when it is applied to a periodic source of gravitational waves at rest; as we show in Sec.~\ref{sec:flux}, an average over a period of oscillation is required to establish the equivalence of the two results. Our new expression allows us to resolve the issue of the Coulombic influence. {\it Contrary to expectations based on the plausibility argument, we find that the flux of angular momentum in general relativity can be written entirely in terms of the radiative degrees of freedom of the gravitational field; Coulombic information, in the form of $M(u=-\infty,\theta,\phi)$ or anything else of the sort, makes no appearance in the flux.} Our conclusion is therefore that nothing was missed, and that the expected analogy between electromagnetic and gravitational fluxes of angular momentum simply breaks down. Why this breakdown occurs is a deep question, for which we currently have no answer.  

Our new expression for the flux of angular momentum in general relativity applies to situations that are more general than those envisioned by Thorne in \cite{thorne}. There is no restriction to sources of gravitational waves that are at rest with respect to the frame in which the flux is evaluated, there is no restriction to periodic sources, and there is no averaging over a period of oscillation. Our new formula applies to all situations involving an asymptotically flat spacetime with a bounded matter source. 

The paper is organized as follows. We introduce the Bondi metric in Sec.~\ref{sec:bondi}, present it in the original coordinates $(u,r,\theta,\phi)$ and in a related Lorentzian system $(t,x,y,z)$, expand it in powers of $r^{-1}$, and write down field equations for the various expansion coefficients. In Sec.~\ref{sec:flux} we introduce the flux ${\scrpt T}^{ab}$ and calculate it with the help of the Bondi metric. We show that the result can be expressed entirely in terms of the field's radiative degrees of freedom, and we compare our expression to Thorne's standard formula. In Sec.~\ref{sec:Jab} we introduce the angular momentum $J^{ab}$, calculate it with the help of the Bondi metric, and verify the validity of the balance law in Eq.~(\ref{eq:balance}). In the course of this discussion we get compelled to alter the definitions slightly, and to introduce alternative notions of angular momentum and flux that are more closely in tune with the Bondi framework. We also address the ambiguity that plagues the definition of angular momentum in general relativity, which is associated with supertranslations, a subgroup of the Bondi-Metzner-Sachs (BMS) group of transformations that preserves the form of the Bondi metric. While the balance equation can be formulated in any Bondi frame, each side of the equation changes (consistently) under a supertranslation. We further explain that in typical situations in which the spacetime is stationary in the remote past, the ambiguity can be eliminated with a choice of preferred Bondi frame. For the sake of completeness, in our final Sec.~\ref{sec:energy} we examine the balance laws for energy and linear momentum in our combined Landau-Lifshitz and Bondi formalisms. 

Several technical developments are relegated to appendices. In Appendix~\ref{sec:super} we calculate the transformation of various quantities under infinitesimal supertranslations, and show that Eq.~(\ref{eq:balance}) is preserved. In Appendix~\ref{app:comp} we provide an explicit listing of components for various tensors that are introduced in Sec.~\ref{sec:flux}. Finally, in Appendix~\ref{app:boosted} we examine the spacetime of a boosted mass, and provide a derivation of Eq.~(\ref{eq:massaspect}). Throughout the paper we use geometrized units and set $G = c = 1$.    

\section{Metric} 
\label{sec:bondi} 

\subsection{Bondi metric and field equations} 

The Bondi metric is an expansion of the metric of an asymptotically flat spacetime in inverse powers of the distance from the matter distribution. It was introduced in \cite{BvBM, Sachs}, and \cite{bondisachsreview} provides a comprehensive review. The metric is presented in coordinates $(u,r,\theta,\phi)$ attached to the spacetime's expanding null cones. The retarded-time coordinate $u$ is constant on each null cone, and the angles $\theta^A = (\theta,\phi)$ are constant on the null generators; the radial coordinate $r$ is a nonaffine parameter on each generator. The definition of the coordinates implies that $g^{uu} = g^{uA} = 0$. The metric functions $U$, $V$, $W^A$ and $\gamma_{AB}$  are defined in terms of the inverse metric; specifically, we have that $g^{ur} = -1/U$, $g^{rr} = V/U$, $g^{rA} = W^A/(rU)$, and $g^{AB} = \gamma^{AB}/r^2$, where $\gamma^{AB}$ is the matrix inverse of $\gamma_{AB}$. The metric is given by 
\begin{equation} 
ds^2 = -UV\, du^2 - 2U\, dudr 
+ \gamma_{AB} ( r\, d\theta^A + W^A\, du ) ( r\, d\theta^B + W^B\, du ). 
\label{bondi_metric} 
\end{equation} 
The scaling of the radial coordinate is fixed by imposing $\mbox{det}[\gamma_{AB}] = \sin^2\theta$. The Minkowski metric is recovered when $U=1, V=1, W^A=0$, and $\gamma_{AB}=\Omega_{AB}$, where $\Omega_{AB} := \mbox{diag}(1,\sin^2\theta)$ is the metric on the unit 2-sphere. 

We assume that the metric is smooth, and that there is no matter outside a bounded region surrounding $r=0$. The metric functions are expressed as asymptotic expansions in powers of $r^{-1}$, and the expansion coefficients are determined by the vacuum field equations. Two functions of $(u,\theta^A)$ are left undetermined by the field equations. These are the radiative  degrees of freedom of the gravitational field, described by  
\begin{equation} 
X(u,\theta^A) := \lim_{r\to\infty} (r h_+), \qquad 
Y(u,\theta^A) := \lim_{r\to\infty} (r h_\times), 
\end{equation} 
where $h_+$ and $h_\times$ are the two polarizations of the gravitational wave. We package these quantities into a symmetric 2-tensor $f_{AB}$ defined on the tangent space of the unit 2-sphere: 
\begin{equation} 
f_{AB} := \left( 
\begin{array}{cc} 
X & Y \sin\theta \\ 
Y \sin\theta & -X \sin^2\theta 
\end{array} \right).
\end{equation} 
The tensor $f_{AB}$ is tracefree, that is $\Omega^{AB} f_{AB} = 0$, where $\Omega^{AB} = \mbox{diag}(1,1/\sin^2\theta)$ is the matrix inverse of $\Omega_{AB}$. We denote the covariant derivative on the unit 2-sphere with the symbol $D_A$, with the understanding that the connection is compatible with $\Omega_{AB}$. We also use the notation 
\begin{equation} 
f^2 := f_{AB} f^{AB} = 2(X^2 + Y^2), 
\end{equation} 
where here and below, an upper-case Latin index such as $A$ is raised with $\Omega^{AB}$; these indices are lowered with $\Omega_{AB}$.  

To the order required to calculate the flux of angular momentum in Sec.~\ref{sec:flux}, the metric functions are expanded as  
\begin{subequations} 
\label{Bondi_expansions} 
\begin{align} 
U &= 1 + B/r^2 + O(r^{-3}), \\ 
V &= 1 - 2M/r + N/r^2 + O(r^{-3}), \\
W^A &= A^A/r + B^A/r^2 + O(r^{-3}), \\ 
\gamma_{AB} &= \Omega_{AB} + f_{AB}/r + \tfrac{1}{4} f^2 \Omega_{AB}/r^2 + O(r^{-3}), \\ 
\gamma^{AB} &= \Omega^{AB} - f^{AB}/r + \tfrac{1}{4} f^2 \Omega^{AB}/r^2 + O(r^{-3}),
\end{align} 
\end{subequations} 
with each coefficient a function of $u$ and $\theta^A$. In the expansions we have invoked the Einstein field equations to eliminate a term at order $r^{-1}$ in $U$, and to determine the term at order $r^{-2}$ in $\gamma_{AB}$. Additionally, the field equations imply that
\begin{equation}
B = -\frac{1}{16} f^2, \qquad 
A^A = \frac{1}{2} D_B f^{AB}, 
\label{bondiFE1} 
\end{equation} 
and up to initial conditions, the Bondi-mass aspect $M(u,\theta^A)$ is determined by  
\begin{equation} 
\frac{\partial M}{\partial u} = -\frac{1}{8} \dot{f}_{AB} \dot{f}^{AB} 
+ \frac{1}{4} D_A D_B \dot{f}^{AB}, 
\label{bondiFE2} 
\end{equation} 
in which $\dot{f}_{AB} := \partial f_{AB}/\partial u$. The celebrated Bondi mass-loss formula follows from Eq.~(\ref{bondiFE2}) after integration over the unit 2-sphere; the first term on the right describes the flux of gravitational-wave energy, and the second term integrates to zero. 

The field equations further imply 
\begin{equation} 
\frac{\partial B^A}{\partial u} = D_B \Gamma^{AB} 
- \frac{1}{2} \dot{f}^A_{\ C} D_B f^{BC}
+ \frac{1}{6} \dot{f}^B_{\ C} D_B f^{AC}, 
\label{bondiFE3a} 
\end{equation} 
where 
\begin{equation} 
\Gamma^{AB} := \Omega^{AB} \biggl( \frac{2}{3} M 
- \frac{1}{16} \frac{\partial f^2}{\partial u} \biggr) 
+ \frac{1}{6} D_C \Bigl( D^A f^{BC} - D^B f^{AC} \Bigr).  
\label{bondiFE3b} 
\end{equation} 
In the literature, $B^A$ is often referred to as an angular-momentum aspect.\footnote{The term ``mass aspect'' was coined in \cite{BvBM}, but the term ``angular-momentum aspect'' was introduced later in \cite{tamburinowinicour}.} Conventions differ, however, on the precise definition of the  angular-momentum aspect, and some authors shift $B^A$ by terms proportional to $f^{AB}$ and its derivatives, and/or multiply it by a numerical factor; see Eq. (2.8) and (2.9) in \cite{Compere:2018ylh} for a comparison between various conventions used in the literature. In Sec.~\ref{sec:balance} we will reveal the link between Eq.~(\ref{bondiFE3a}) and the statement of conservation of angular momentum. 

An equation can also be written down for the remaining function $N$, but this will not be required in our further  developments.

\subsection{Transformation to Lorentzian coordinates} 
\label{sec:transf} 

We transform the Bondi metric from the coordinates $(u,r,\theta^A)$ to a Lorentzian system $(t,x,y,z)$ defined by  
\begin{equation} 
t = u + r, \qquad 
x^a = r \; \Omega^a(\theta^A), 
\end{equation} 
in which $x^a := (x,y,z)$ and $\Omega^a := (\sin\theta\cos\phi,\sin\theta\sin\phi,\cos\theta)$.
We introduce the notations 
\begin{equation} 
\Omega^a_A := \partial_A \Omega^a, \qquad 
\Omega^A_a := \Omega^{AB} \delta_{ab} \Omega^b_B, 
\end{equation} 
and note that these objects satisfy the identities 
\begin{equation} 
\Omega_a \Omega^a_A = 0, \qquad 
\Omega_{AB} = \delta_{ab} \Omega^a_A \Omega^b_B, \qquad 
\Omega^{AB} \Omega^a_A \Omega^b_B = \delta^{ab} - \Omega^a \Omega^b, \qquad 
D_A D_B \Omega^a = -\Omega^a \Omega_{AB}.  
\label{Oidentities} 
\end{equation}   
We use the Euclidean metric $\delta_{ab}$ to lower and raise all lower-case Latin indices.

When transforming a 2-tensor like $f_{AB}$ to Cartesian coordinates, it is convenient to omit factors of $r$ and work entirely on the unit 2-sphere. One must then take care to re-insert the factors of $r$ at an appropriate later stage. We therefore introduce the notation 
\begin{equation} 
A^a := \Omega^a_A\, A^A, \qquad 
f^{ab} := \Omega^a_A \Omega^b_B\, f^{AB}, 
\end{equation} 
and so on. In a similar fashion we let 
\begin{equation} 
f^{ab}_{\ \ |c} :=  \Omega^a_A \Omega^b_B \Omega^C_c\, D_C f^{AB}, 
\end{equation} 
and so forth. In this notation, the second of Eqs.~(\ref{bondiFE1}) becomes $A^a = \frac{1}{2} f^{ab}_{\ \ |b}$, and we also note that $f^2 = f_{ab} f^{ab}$. 

A straightforward calculation reveals that the inverse metric becomes
\begin{equation} 
g^{tt} = \frac{1}{U}(V-2), \qquad 
g^{ta} = \frac{1}{U} \bigl[ (V-1) \Omega^a + W^a \bigr], \qquad 
g^{ab} = \frac{1}{U} \bigl( V \Omega^a \Omega^b + \Omega^a W^b + W^a \Omega^b + U \gamma^{ab} \bigr) 
\label{g_inv_exact} 
\end{equation} 
in the Lorentzian coordinates. We have that $\sqrt{-g} = U$, and the components of $\gothg^{\alpha\beta} := \sqrt{-g} g^{\alpha\beta}$ are 
\begin{equation} 
\gothg^{tt} = V-2, \qquad 
\gothg^{ta} = (V-1) \Omega^a + W^a, \qquad 
\gothg^{ab} = V \Omega^a \Omega^b + \Omega^a W^b + W^a \Omega^b + U \gamma^{ab}.  
\label{gothg_exact}  
\end{equation} 
Taking into account the expansions in powers of $1/r$, we obtain 
\begin{subequations} 
\label{gothg} 
\begin{align} 
\gothg^{tt} &= -1 - \frac{2M}{r} + \frac{N}{r^2} + O(r^{-3}), \\
\gothg^{ta} &= \bigl( -2M \Omega^a + A^a \bigr) \frac{1}{r} 
+ \bigl( N \Omega^a + B^a \bigr) \frac{1}{r^2} + O(r^{-3}), \\
\gothg^{ab} &= \delta^{ab} + \bigl( -2M\Omega^a \Omega^b + \Omega^a A^b + A^a \Omega^b 
- f^{ab} \bigr) \frac{1}{r} 
\nonumber \\ & \quad \mbox{} 
+ \Bigl[ N \Omega^a \Omega^b 
+ \Omega^a B^b + B^a \Omega^b + \tfrac{3}{16} f^2 \bigl( \delta^{ab} - \Omega^a \Omega^b \bigr) 
\Bigr] \frac{1}{r^2} + O(r^{-3}). 
\end{align} 
\end{subequations}
To the order required in Sec.~\ref{sec:flux}, the components of the inverse metric are 
\begin{subequations} 
\label{g_inverse} 
\begin{align} 
g^{tt} &= -1 - \frac{2M}{r} + O(r^{-2}), \\
g^{ta} &= \bigl( -2M \Omega^a + A^a \bigr) \frac{1}{r} + O(r^{-2}), \\
g^{ab} &= \delta^{ab} + \bigl( -2M\Omega^a \Omega^b + \Omega^a A^b + A^a \Omega^b 
- f^{ab} \bigr) \frac{1}{r} + O(r^{-2}), 
\end{align} 
\end{subequations} 
and those of the metric are 
\begin{subequations} 
\label{g} 
\begin{align} 
g_{tt} &= -1 + \frac{2M}{r} + O(r^{-2}), \\
g_{ta} &= \bigl( -2M \Omega_a + A_a \bigr) \frac{1}{r} + O(r^{-2}), \\
g_{ab} &= \delta_{ab} + \bigl( 2M\Omega_a \Omega_b - \Omega_a A_b - A_a \Omega_b 
+ f_{ab} \bigr) \frac{1}{r} + O(r^{-2}). 
\end{align} 
\end{subequations} 
As stated previously, all lower-case Latin indices on the right-hand side of these equations are lowered with $\delta_{ab}$. 

The expansion coefficients $M$, $f_{ab}$, $A^a$, and so on, are still viewed as functions of $u$ and $\theta^A$, and the Lorentzian components of $\gothg^{\alpha\beta}$ are therefore viewed as functions of $u$, $r$, and $\theta^A$. The partial-derivative operator $\partial_\gamma$, however, refers to the Lorentzian coordinates, and derivatives of $\gothg^{\alpha\beta}$ are calculated as 
\begin{equation} 
\partial_\gamma \gothg^{\alpha\beta} = 
\frac{\partial \gothg^{\alpha\beta}}{\partial u}\, \partial_\gamma u 
+ \frac{\partial \gothg^{\alpha\beta}}{\partial r}\, \partial_\gamma r 
+ \frac{\partial \gothg^{\alpha\beta}}{\partial \theta^A}\, \partial_\gamma \theta^A. 
\end{equation} 
We have that $\partial_t u = 1$, $\partial_a u = -\Omega_a$, $\partial_t r = 0$, $\partial_a r = \Omega_a$, $\partial_t \theta^A = 0$, and $\partial_a \theta^A = r^{-1} \Omega^A_a$. 

\section{Flux of angular momentum} 
\label{sec:flux} 

The flux of angular momentum is calculated in the Landau-Lifshitz formalism, without specializing to harmonic coordinates. The densitized inverse metric of Eqs.~(\ref{gothg}), the inverse metric of Eqs.~(\ref{g_inverse}), and the metric of Eqs.~(\ref{g}) are inserted into the Landau-Lifshitz pseudotensor $t^{\alpha\beta} := (-g) t^{\alpha\beta}_{\rm LL}$ (see Eq.~(6.5) of \cite{pw}), which is expanded through orders $r^{-2}$ and $r^{-3}$. This is then substituted into the expression for the flux of angular momentum (see Eq.~(12.36) of \cite{pw}),  
\begin{equation} 
{\scrpt T}^{ab} = \oint \bigl( x^a t^{bc} - x^b t^{ac} \bigr)\, dS_c.  
\end{equation} 
The integral is evaluated in the limit $r \to \infty$, and $dS_c = \Omega_c r^2\, d\Omega$ (with $d\Omega := \sin\theta\, d\theta d\phi$) is the surface element on a coordinate sphere of constant $u$ and $r$. 
After a rather long calculation, we obtain 
\begin{equation} 
{\scrpt T}^{ab} = \int \gotht^{ab}\, d\Omega,
\label{flux1} 
\end{equation} 
where 
\begin{equation} 
\gotht^{ab} = -\frac{1}{16\pi} \Omega^{[a} \bigl( -2 f^{b]}_{\ c}\, \dot{A}^c + 2 \dot{f}^{b]}_{\ c}\, A^c 
+ \dot{f}^{b]}_{\ c}\, f^{cd}_{\ \ |d} - f^{b]}_{\ c|d}\, \dot{f}^{cd} \bigr). 
\end{equation} 
A truly remarkable aspect of this result is that the flux integrand $\gotht^{ab}$ depends only on the metric functions $f^{ab}$ and $A^a$; it is independent of the mass aspect $M$ and all other functions that appear in the metric at the relevant orders in $r^{-1}$. We have verified that these quantities also do not appear in the Landau-Lifshitz pseudotensor expanded through order $r^{-3}$.  

Incorporating the field equation $A^a = \frac{1}{2} f^{ab}_{\ \ |b}$ and isolating a total $u$-derivative, an equivalent expression for the flux integrand is 
\begin{equation} 
\gotht^{ab} = \frac{\partial \gothr^{ab}}{\partial u} + \goths^{ab}, 
\label{q_final} 
\end{equation} 
where 
\begin{equation} 
\gothr^{ab} := 
\frac{1}{16\pi} \Omega^{[a} f^{b]}_{\ c}\, f^{cd}_{\ \ |d} 
= \frac{1}{16\pi} \Omega^{[a} \Omega^{b]}_B f^{B}_{\ C}\, D_D f^{CD}  
\label{eq:rab}
\end{equation} 
and 
\begin{equation} 
\goths^{ab} := -\frac{1}{16\pi} \Omega^{[a} \bigl( 3 \dot{f}^{b]}_{\ c}\, f^{cd}_{\ \ |d} 
- f^{b]}_{\ c|d}\, \dot{f}^{cd} \bigr) 
= - \frac{1}{16\pi} \Omega^{[a} \Omega^{b]}_B \bigl( 3 \dot{f}^{B}_{\ C}\, D_D f^{CD} 
- \dot{f}^{CD}\, D_C f^B_{\ D} \bigr). 
\label{eq:sab}
\end{equation} 

The ``standard expression'' for the flux integrand, used everywhere in the literature on gravitational waves, is given by 
\begin{subequations}
\label{q_standard} 
\begin{align}  
\gotht^{ab}_{\rm standard} &= 
-\frac{1}{16\pi} \Omega^{[a} \bigl( \dot{f}^{b]}_{\ c}\, f^{cd}_{\ \ |d}  
+ f^{b]}_{\ c|d}\, \dot{f}^{cd} \bigr) 
+ \frac{1}{8\pi} f^{[a}_{\ \ c}\, \dot{f}^{b]c} \\ 
&= -\frac{1}{16\pi} \Omega^{[a} \Omega^{b]}_B \bigl( \dot{f}^B_{\ C}\, D_D f^{CD} 
+ \dot{f}^{CD}\, D_C f^B_{\ D} \bigr) 
+ \frac{1}{8\pi} \Omega^{[a}_A \Omega^{b]}_B f^A_{\ C}\, \dot{f}^{CB}. 
\end{align}
\end{subequations}
The standard expression originates in Sec.~IV D of \cite{thorne}. In the discussion, Thorne makes it clear that Eq.~(\ref{q_standard}) is meant to apply only to sources of gravitational waves that are at rest relative to the reference frame in which the flux is evaluated. Furthermore, the formula is meant to apply only to periodic sources of gravitational waves, and it involves an average over a period of oscillation. 

The standard expression differs from our own expression for the flux of angular momentum. The difference between Eqs.~(\ref{q_final}) and (\ref{q_standard}) is  
\begin{equation} 
\gotht^{ab} - \gotht^{ab}_{\rm standard} = \frac{\partial \gothp^{ab}}{\partial u} 
- D_D \biggl( \frac{1}{8\pi} \Omega^{[a} \Omega^{b]}_B \dot{f}^B_{\ C}\, f^{CD} \biggr), 
\end{equation} 
where 
\begin{equation} 
\gothp^{ab} := 
\frac{1}{16\pi} \Omega^{[a} \bigl( f^{b]}_{\ c}\, f^{cd}_{\ \ |d} 
+ 2 f^{b]}_{\ c|d} f^{cd} \bigr) 
= \frac{1}{16\pi} \Omega^{[a} \Omega^{b]}_B \bigl( f^B_{\ C}\, D_D f^{CD} 
+ 2 f^{CD}\, D_C f^B_{\ D} \bigr). 
\end{equation} 
The second term of this difference is a divergence on the unit 2-sphere, and its integral over the angles $(\theta,\phi)$ vanishes. The first term is a total time derivative, which necessarily vanishes when averaged over a period of oscillation. We conclude that Eq.~(\ref{q_final}) and the standard expression are equivalent (after averaging) when applied to a periodic source of gravitational waves.   

In contrast to the standard expression, Eq.~(\ref{q_final}) is not restricted to periodic sources of gravitational waves, and there is no requirement that the source be at rest. Its domain of applicability is therefore wider than the standard expression. We note also that our derivation did not rely on a post-Minkowskian expansion of the metric; our result is therefore valid to all orders in $G$. It holds in full general relativity, for any asymptotically flat spacetime. 

An explicit listing of components for the tensors $\gothr^{ab}, \goths^{ab}, \gotht^{ab}_{\rm standard}$, and $\gothp^{ab}$ is provided in Appendix~\ref{app:comp}. The calculations in this section were carried out with the help of GRTensorIII \cite{grtensor} working under Maple and, independently, in Mathematica with the Riemannian Geometry and Tensor Calculus package. 

\section{Angular momentum and balance law} 
\label{sec:Jab}

\subsection{Angular momentum} 

In the Landau-Lifshitz formalism, the angular-momentum tensor is defined as an integral over a 2-sphere $(u,r) = \mbox{constant}$ in the limit $r \to \infty$ (see Eq.(6.25) in \cite{pw}). We have that 
\begin{align}
J^{ab} & = \frac{1}{8 \pi} \int \left( r \; \Omega^{[a} \del_\mu H^{b]\mu t c} \; \Omega_c - H^{t[ab]k} \Omega_k \right) r^2 \; \d \Omega , 
\label{eq:def-Jab}
\end{align}
where $H^{\mu \nu \kappa \lambda} := 2\gothg^{\mu [\kappa} \gothg^{ \lambda]\nu}$, with $\gothg^{\mu \nu}:= \sqrt{-g} g^{\mu\nu}$. We write this as 
\begin{equation} 
J^{ab} = \int \gothj^{ab}\, d\Omega, 
\end{equation} 
and refer to $\gothj^{ab}$ as the angular-momentum integrand. Inserting Eqs.~(\ref{gothg_exact}) for $\gothg^{\mu \nu}$, we find that
\begin{equation} 
 \gothj^{ab} = \frac{r^2}{4\pi} \left[ - U (\gamma^c_{\ c} - \gamma_{cd} \Omega^c \Omega^d) \Omega^{[a} W^{b]} + 
 r \Omega^{[a} \partial_r W^{b]} + U \Omega^{[a} \gamma^{b]c} W_c \right]
 \label{eq:angmom-integrand}
\end{equation} 
where $\partial_r$ is a partial derivative with respect to $r$ that leaves $u$ alone, in spite of the relation $u = t - r$.  
This expression makes it clear that the leading-order part of $\gothj^{ab}$ in the limit $r \to \infty$ is sensitive to terms of order $r^{-3}$ in the asymptotic expansion of the metric. 

The expansions of Eq.~(\ref{Bondi_expansions}) must therefore be extended to include additional terms that were not required in the computation of $\gotht^{ab}$. We write 
\begin{subequations} 
\label{Bondi_expansions2} 
\begin{align} 
U &= 1 + B/r^2 + C/r^3 + O(r^{-4}), \\ 
W^A &= A^A/r + B^A/r^2 + C^A/r^3 + O(r^{-4}), \\ 
\gamma^{AB} &= \Omega^{AB} - f^{AB}/r + \tfrac{1}{4} f^2 \Omega^{AB}/r^2 
+ h^{AB}/r^3 + O(r^{-4}), 
\end{align} 
\end{subequations} 
and note that $h^{AB}$ must be tracefree, $\Omega_{AB} h^{AB} = 0$, to ensure that $\mbox{det}[\gamma^{AB}] = 1/\sin^2\theta$. Substituting these expansions into Eq.~(\ref{eq:angmom-integrand}), we find that  
\begin{equation} 
\gothj^{ab} = -\frac{1}{4\pi} r \Omega^{[a} A^{b]} 
- \frac{1}{8\pi} \Omega^{[a} \bigl( 3 B^{b]} + f^{b]}_{\ c} A^c \bigr) + O(r^{-1}).  
\end{equation} 
Remarkably, the result turns out to be independent of all $O(r^{-3})$ terms in the metric, and indeed, of most metric functions at order $r^{-1}$ and $r^{-2}$; the only relevant ingredients are $f^{ab}$ and $A^a = \frac{1}{2} f^{ab}_{\ \ |b}$, which occur at order $r^{-1}$, and $B^a$, which occurs at order $r^{-2}$.

The angular momentum is to be evaluated in the limit $r \to \infty$, and we observe that the first term in $\gothj^{ab}$ diverges in this limit. This term, however, vanishes after angular integration. To see this, we write 
\begin{equation} 
\Omega^a A^b = \Omega^a \Omega^b_B A^B 
= \frac{1}{2} \Omega^a \Omega^b_B D_C f^{BC} 
= \frac{1}{2} D_C \bigl( \Omega^a \Omega^b_B f^{BC} \bigr) 
- \frac{1}{2} \Omega^a_A \Omega^b_B f^{AB}, 
\end{equation} 
and observe that the second term vanishes after antisymmetrization with respect to $a$ and $b$, and that the divergence produces a vanishing contribution to the angular momentum after integration. We may therefore discard the term of order $r$ in $\gothj^{ab}$, take the limit $r \to \infty$, and thereby obtain a regularized angular-momentum integrand, 
\begin{align} 
\gothj^{ab}_{\rm reg} &= -\frac{1}{16\pi} \Omega^{[a} \bigl( 6 B^{b]} + f^{b]}_{\ c} f^{cd}_{\ \ |d}\bigr) 
\nonumber \\ 
&= -\frac{1}{16\pi} \Omega^{[a} \Omega^{b]}_B \bigl( 6 B^{B} + f^{B}_{\ C} D_D f^{CD} \bigr) 
\nonumber \\ 
&= -\frac{3}{8\pi} \Omega^{[a} B^{b]} - \gothr^{ab}. 
\label{j_final} 
\end{align}
In contrast to the flux of angular momentum, which is entirely determined by the radiative aspects of the gravitational field, 
$\gothj^{ab}_{\rm reg}$ depends on both radiative and ``Coulombic'' aspects of the field, with the Coulombic aspects 
encoded in $B^a$. In Appendix~\ref{app:boosted} we show that $\gothj^{ab}_{\rm reg}$ is nonzero for a spacetime describing a boosted  black hole, but that the angular momentum $J^{ab}$ properly vanishes.   

Our expression $J^{ab} = \int j^{ab}_{\rm reg}\, d\Omega$ for the total angular momentum does not agree with analogous results that can be found in the literature \cite{DrayStreubel,WaldZoupas,FlanaganNichols}.
Because other authors adopt different {\it definitions} for angular momentum, for example by utilizing spinor or covariant phase-space methods, there is no reason why these expressions should all agree.%
\footnote{Nonetheless, these different definitions all agree in non-radiative regions of future null infinity, in a canonical Bondi frame.}
Our own definition is grounded in the Landau-Lifshitz formalism, which, as we explained back in Sec.~\ref{sec:intro}, supplies us with a precise, sound, convenient, and widely used notion of angular momentum, together with a balance law that allows us to calculate its flux.      

\subsection{Balance law} 
\label{sec:balance} 

With expressions in hand for the angular-momentum flux and total angular momentum, we may return to the statement of angular momentum balance, 
\begin{equation} 
\frac{d}{du} J^{ab} = -{\scrpt T}^{ab}, 
\end{equation} 
or, equivalently,
\begin{equation} 
\frac{d}{du} \int \gothj^{ab}_{\rm reg}\, d\Omega 
= -\int \gotht^{ab}\, d\Omega. 
\end{equation} 
We see that the $\gothr^{ab}$ term in $\gothj^{ab}_{\rm reg}$ is a perfect match for the $\gothr^{ab}$ term in $\gotht^{ab}$, as given by Eq.~(\ref{q_final}). These terms, therefore, can simply be removed from the statement of angular-momentum conservation. Introducing the new definitions 
\begin{equation} 
{\sf J}^{ab} := \int {\sf j}^{ab}\, d\Omega, \qquad 
{\sf j}^{ab} := -\frac{3}{8\pi} \Omega^{[a} B^{b]}  
\label{eq:Jnewdef}
\end{equation} 
and 
\begin{equation} 
{\sf T}^{ab} := \int {\sf t}^{ab}\, d\Omega, \qquad  
{\sf t}^{ab} := \goths^{ab} = -\frac{1}{16\pi} \Omega^{[a} \bigl( 3 \dot{f}^{b]}_{\ c}\, f^{cd}_{\ \ |d} 
- f^{b]}_{\ c|d}\, \dot{f}^{cd} \bigr),  
\label{eq:Tnewdef}
\end{equation} 
the balance law can be expressed as $d{\sf J}^{ab}/du = -{\sf T}^{ab}$. Note that ${\sf j}^{ab}$ is entirely determined by $B^a$,
and this provides justification for the interpretation of $B^a$ as an angular-momentum aspect. The new definition of angular momentum is closely associated with the Bondi formalism, but we see that it differs very little from the Landau-Lifshitz definition. 

The statement of angular-momentum balance, as redefined here, can also be obtained directly from the vacuum field equations. Indeed, Eqs.~(\ref{bondiFE3a}) and (\ref{bondiFE3b}) imply that 
\begin{equation} 
\frac{\partial {\sf j}^{ab}}{\partial u} = -\frac{3}{8\pi} \Omega^{[a} \Omega^{b]}_B D_C \Gamma^{BC}
- {\sf t}^{ab}. 
\end{equation} 
The identity 
\begin{equation} 
\Omega^{[a} \Omega^{b]}_B D_C \Gamma^{BC} = D_C \biggl( 
\Omega^{[a} \Omega^{b]}_B \Gamma^{BC} 
+ \frac{1}{3} \Omega^{[a}_A \Omega^{b]}_B D^A f^{BC} 
+ \frac{1}{3} \Omega^{[a} \Omega^{b]}_B f^{BC} \biggr) 
\end{equation} 
reveals that the first term vanishes after angular integration, and we again arrive at $d{\sf J}^{ab}/du = -{\sf T}^{ab}$.  

\subsection{Ambiguity under supertranslations} 
\label{sec:canonicalframe}

It is well known that the angular momentum of a radiative spacetime is ambiguous, thanks to the Bondi-Metzner-Sachs (BMS) group of transformations that preserve the form of the Bondi metric \cite{Sachs-BMS,DrayStreubel,Nichols}. The BMS group contains a subgroup of rotations and boosts, and another subgroup of supertranslations, which consist of angle-dependent translations of the retarded-time coordinate accompanied by angle-dependent translations of the spatial coordinates. Because they are Cartesian tensors, the transformation of ${\sf j}^{ab}$ and ${\sf t}^{ab}$ under rotations and boosts is well understood, and there is no need to consider them here. Our concern is instead with the supertranslations, which are at the source of the ambiguity that plagues angular momentum. In Appendix~\ref{sec:super} we show that 
${\sf j}^{ab}$ and ${\sf t}^{ab}$ do change under an infinitesimal supertranslation; this implies that the angular momentum ${\sf J}^{ab}$ and flux ${\sf T}^{ab}$ both depend on a choice of Bondi frame. We also show, however, that these quantities change in a consistent way, such that the balance law 
$d {\sf J}^{ab}/du = -{\sf T}^{ab}$ is always preserved; the balance law is therefore valid in any Bondi frame.  

Under the typical assumption that the spacetime is stationary in the remote past, it is possible to remove the supertranslation ambiguity by adopting a preferred Bondi frame. A prescription to achieve this is nicely described in Sec.~II~D of \cite{FlanaganNichols}; it dates back to earlier work by van der Burg and Bondi \cite{BurgBondi} and Newman and Penrose \cite{NewmanPenrose}. The key element is to require that $f_{AB}$ vanishes in the remote past (instead of being merely constant). This means that the gravitational-wave polarizations are set to zero before any dynamical process takes place. As Eq.~(\ref{deltaf}) shows, a supertranslation would turn an initially zero $f_{AB}$ into a non-zero $f_{AB}(\theta,\phi)$, and this would produce a violation of the requirement. In this way the freedom to perform supertranslations is eliminated, and the gauge is completely fixed. In the preferred Bondi frame, ${\sf J}^{ab}$ and ${\sf T}^{ab}$ are unambiguous.  
 
By requiring that $f_{AB} = 0$ in the remote past, the gravitational-wave polarizations --- suitably identified as the transverse-tracefree part of the metric at future null infinity --- become gauge-invariant, and they can therefore be computed in any coordinate system; there is no necessity to rely exclusively on Bondi coordinates. For example, in a typical application the waveforms could be obtained to a desired post-Newtonian order by working in harmonic coordinates, provided that sufficient care is taken to express them in terms of the correct retarded-time variable (the one that is constant on outgoing null cones of the post-Newtonian spacetime \cite{isaacson-winicour:68, blanchet:87}).  
In another application, the waveforms could be obtained from the Weyl scalar $\Psi_4$ by integrating the Teukolsky equation in the background of a Kerr spacetime. And once the gauge-invariant waveforms are at hand, they can be freely inserted within our expression for the angular-momentum flux.   

\section{Energy and linear momentum} 
\label{sec:energy} 

In this section we examine, for the sake of completeness, the statements of energy and linear-momentum balance in the Landau-Lifshitz formalism. We once more rely on the Bondi metric to calculate all quantities that appear in these equations. We show that energy balance takes the form of the celebrated Bondi mass-loss formula, and we recover the familiar expression of momentum balance. 

Our starting point is the Landau-Lifshitz statements of energy and linear-momentum balance, as given by Eqs.~(12.31)--(12.34) in \cite{pw}. For energy we have 
\begin{equation} 
\frac{dE}{du} = -{\scrpt P} = -\int \gothp\, d\Omega, \qquad 
\gothp := r^2 t^{0b}\Omega_b, 
\end{equation} 
where $E = \int \gothe\, d\Omega$ is defined by Eq.~(6.36b) in \cite{pw}. For momentum we have   
\begin{equation} 
\frac{dP^a}{du} = -{\scrpt F}^a = -\int \gothf^a\, d\Omega, \qquad 
\gothf^a := r^2 t^{ab}\Omega_b, 
\end{equation} 
with $P^a = \int \gothp^a\, d\Omega$ defined by Eq.~(6.37b) in \cite{pw}. We recall that $t^{\alpha\beta} := (-g) t^{\alpha\beta}_{\rm LL}$ is the Landau-Lifshitz pseudotensor. 

Involving the Bondi metric and calculating as previously, we find that
\begin{equation} 
\gothp = \frac{1}{32\pi} \dot{f}_{AB} \dot{f}^{AB}, \qquad 
\gothf^a = \frac{1}{32\pi} \dot{f}_{AB} \dot{f}^{AB}\, \Omega^a. 
\end{equation} 
These reproduce the standard expressions for energy and momentum fluxes, as listed in Eqs.~(12.45) and (12.46) of \cite{pw}. We also find that the energy and momentum integrands are 
\begin{equation} 
\gothe = \frac{1}{4\pi} M - \frac{1}{32\pi} D_A D_B f^{AB}, \qquad 
\gothp^a = \frac{1}{4\pi} M \Omega^a - \frac{1}{32\pi}D_A \bigl( \Omega^a D_B f^{AB} \bigr).  \label{eq:e-and-mom}
\end{equation} 

The balance equations can be derived directly from the Einstein field equations. Taking the first derivative of the first of 
Eqs.~(\ref{eq:e-and-mom}) and inserting Eq.~(\ref{bondiFE2}), we obtain 
\begin{equation} 
\frac{\partial \gothe}{\partial u} = -\frac{1}{32\pi} \dot{f}_{AB} \dot{f}^{AB}
+ \frac{1}{32\pi} D_A D_B \dot{f}^{AB}; 
\end{equation} 
angular integration then returns $dE/du = -{\scrpt P}$. Similarly, we find that 
\begin{equation} 
\frac{\partial \gothp^a}{\partial u} = -\frac{1}{32\pi} \dot{f}_{AB} \dot{f}^{AB}\, \Omega^a 
+ \frac{1}{16\pi} \bigl( D_A D_B \dot{f}^{AB} \bigr) \Omega^a 
- \frac{1}{32\pi} D_A \bigl( \Omega^a D_B \dot{f}^{AB} \bigr), 
\end{equation}
which can be written in the alternative form 
\begin{equation} 
\frac{\partial \gothp^a}{\partial u} = -\frac{1}{32\pi} \dot{f}_{AB} \dot{f}^{AB}\, \Omega^a 
+ \frac{1}{32\pi} D_A \bigl( \Omega^a D_B \dot{f}^{AB} \bigr)
- \frac{1}{16\pi} D_B \bigl( \Omega^a_A \dot{f}^{AB} \bigr).  
\end{equation}
Angular integration yields $dP^a/du = -{\scrpt F}^a$.

\section*{Acknowledgements}
We express our sincere gratitude to an anonymous referee, who pointed out a serious error in a previous version of this paper. Our original analysis was carried out in harmonic coordinates instead of Bondi coordinates, but it erroneously omitted $\ln r$ terms in the asymptotic expansion of the metric. (We mistakenly thought that such terms could trivially be re-incorporated at the end of the calculation, through a redefinition of the retarded-time coordinate.) As a consequence of this error, we incorrectly found that Coulombic terms did appear in the flux of angular momentum, in contradiction with the correct conclusion of this paper. We also thank Huan Yang and Abhay Ashtekar for many fruitful discussions. BB is also grateful to Jeffrey Winicour for email exchanges. This research was supported in part by Perimeter Institute for Theoretical Physics; research at Perimeter Institute is supported by the Government of Canada through the Department of Innovation, Science and Economic Development Canada and by the Province of Ontario through the Ministry of Research, Innovation and Science. The research was also supported by the Natural Sciences and Engineering Council of Canada. 


\appendix

\section{Supertranslations} 
\label{sec:super} 

The Bondi-Metzner-Sachs (BMS) group of transformations \cite{Sachs-BMS} that preserve the Bondi form of the metric includes rotations, boosts and supertranslations. In this appendix we examine the effect of an infinitesimal supertranslation on the angular momentum ${\sf J}^{ab}$ and the flux of angular momentum ${\sf T}^{ab}$. We show explicitly that the balance equation $d{\sf J}^{ab}/du = -{\sf T}^{ab}$ is preserved by a supertranslation; both sides of the equation change in a consistent manner. 

\subsection{Infinitesimal supertranslation} 
An infinitesimal supertranslation is characterized by an arbitrary function $\alpha(\theta^A)$, which represents an angle-dependent translation of the retarded-time coordinate, accompanied by an angle-dependent translation of the spatial coordinates. The transformation is generated by the vector $\xi^\alpha$ with components 
\begin{subequations} 
	\begin{align} 
	\xi^u &= \alpha, \\
	\xi^r &= \frac{1}{2} D^2\alpha - \frac{1}{4r} \Bigl[ 2 \alpha_A D_B f^{AB} 
	+ f^{AB} D_A \alpha_B \Bigr] + O(r^{-2}), \\ 
	\xi^A &= -\frac{1}{r} \alpha^A + \frac{1}{2r^2} f^{AB} \alpha_B 
	- \frac{1}{16r^3} f^2 \alpha^A + O(r^{-4}), 
	\end{align} 
\end{subequations} 
where $\alpha_A := D_A \alpha$, $D^2 \alpha := D^A \alpha_A$, and $f^2 := f_{AB} f^{AB}$.

The infinitesimal supertranslation produces a change in the metric (and its inverse). 
A straightforward calculation reveals the corresponding changes in the metric functions $f^{AB}$, $M$, and $B^A$. We obtain \footnote{We note that our expression for $\delta B^A$ does not agree with Eq.~(2.18c) of \cite{FlanaganNichols}, even when we account for different notations and specialize their expression to a pure supertranslation.} 
\begin{subequations} 
	\label{deltaBondi} 
	\begin{align} 
	\delta f^{AB} &= \alpha \dot{f}^{AB} - 2 D^A \alpha^B + \Omega^{AB} D^2\alpha, 
	\label{deltaf} \\ 
	\delta M &= \alpha \dot{M} + \frac{1}{4} \dot{f}^{AB} D_A \alpha_B + \frac{1}{2} \alpha_A D_B \dot{f}^{AB}, \\ 
	\delta B^A &= \alpha \dot{B}^A + 2M \alpha^A + f^{AB} \alpha_B + \frac{1}{2} f^{AB} D_B D^2\alpha + \frac{1}{4} D^A \bigl( f^{BC} D_B \alpha_C \bigr) + \frac{1}{2} D^A \bigl( D_C f^{BC}  \alpha_B \bigr)
	\nonumber \\ & \quad \mbox{} 
	+\frac{1}{2} D_C f^{BC} D_B \alpha^A  - \frac{1}{2} D^B \bigl( D_C f^{AC}  \alpha_B \bigr) - \frac{1}{16} \frac{\partial f^2}{\partial u} \alpha^A.
	\label{deltaB} 
	\end{align} 
\end{subequations}   
We recall that an overdot indicates differentiation with respect to $u$. 

\subsection{Change in flux} 
In the definition of Eq.~(\ref{eq:Tnewdef}), the flux of angular momentum is given by  
\begin{equation} 
{\sf T}^{ab} := \int {\sf t}^{ab}\, d\Omega, \qquad 
{\sf t}^{ab} := -\frac{1}{16\pi} \Omega^{[a} \Omega^{b]}_A\, s^A, 
\end{equation} 
where 
\begin{equation} 
s^A := 3 \dot{f}^A_{\ B} D_C f^{BC} - \dot{f}^{BC} D_B f^A_{\ C}. 
\label{sA_def} 
\end{equation} 
We wish to calculate $\delta {\sf t}^{ab}$, the change in ${\sf t}^{ab}$ under an infinitesimal supertranslation. 
To obtain $\delta s^A$ we make use of Eq.~(\ref{deltaf}), and simplify the result by permuting covariant derivatives with the Riemann tensor $R^A_{\ BCD} = \delta^A_{\ C} \Omega_{BD} - \delta^A_{\ D} \Omega_{BC}$ and using the identity $\dot{f}^A_{\ C} \dot{f}^{C B} = \frac{1}{2} (\dot{f}_{CD} \dot{f}^{CD}) \Omega^{AB}$, which applies to any symmetric-tracefree tensor. After a straightforward computation we find that 
\begin{equation} 
\delta s^A = \alpha \dot{s}^A + \dot{f}_{BC} \dot{f}^{BC} \alpha^A - 8 \dot{f}^{AB} \alpha_B
-4 \dot{f}^{AB} D_B D^2 \alpha + 2 \dot{f}^{BC} D^A D_B \alpha_C. 
\end{equation} 
It then follows that 
\begin{equation} 
-16\pi \delta {\sf t}^{ab} = \Omega^{[a} \Omega^{b]}_A \bigl( \alpha \dot{s}^A + \dot{f}_{BC} \dot{f}^{BC} \alpha^A 
- 8 \dot{f}^{AB} \alpha_B -4\dot{f}^{AB} D_B D^2 \alpha 
+ 2 \dot{f}^{BC} D^A D_B \alpha_C \bigr), 
\label{delta_t} 
\end{equation} 
and $\delta {\sf T}^{ab}$ is obtained after integration over the unit two-sphere.     

\subsection{Change in angular momentum} 

In the definition of Eq.~(\ref{eq:Jnewdef}), the total angular momentum is given by  
\begin{equation} 
{\sf J}^{ab} := \int {\sf j}^{ab}\, d\Omega, \qquad 
{\sf j}^{ab} := -\frac{1}{16\pi} \Omega^{[a} \Omega^{b]}_A (6 B^A), 
\end{equation} 
and we wish to calculate the change in ${\sf j}^{ab}$ under an infinitesimal supertranslation. Because $\delta {\sf j}^{ab}$ is to be integrated over the unit two-sphere to yield $\delta {\sf J}^{ab}$, we shall calculate it up to terms that vanish upon integration. For this purpose it is useful to introduce the following notion of equivalence: Two quantities $A^{ab}(\theta^A)$ and $B^{ab}(\theta^A)$ shall be declared equivalent when they differ by a third quantity that vanishes after angular integration. In symbols, we write $A^{ab} \sim B^{ab}$ when $A^{ab} = B^{ab} + D_A C^{abA}$ for some $C^{abA}(\theta^A)$.

Combining Eqs.~(\ref{bondiFE3a}) and (\ref{deltaB}), we have that 
\begin{align} 
6 \delta B^A &= \alpha D_B (6 \Gamma^{AB}) + \alpha \bigl( -3 \dot{f}^A_{\ C} D_B f^{BC} + \dot{f}^B_{\ C} D_B f^{AC} \bigr) 
+ 12 M \alpha^A + 6f^{AB} \alpha_B + 3 f^{AB} D_B D^2\alpha 
+ \frac{3}{2} D^A \bigl( f^{BC} D_B \alpha_C \bigr) 
\nonumber \\ & \quad \mbox{} 
+ 3 D^A \bigl( D_C f^{BC}  \alpha_B \bigr)
+ 3 D_C f^{BC} D_B \alpha^A  
-3 D^B \bigl( D_C f^{AC}  \alpha_B \bigr)
- \frac{3}{8} \frac{\partial f^2}{\partial u} \alpha^A, 
\end{align} 
and we obtain $\delta {\sf j}^{ab}$ by multiplying this by $\Omega^{[a} \Omega^{b]}_A$. We shall simplify the resulting expression by exploiting the notion of equivalence; our general strategy is to let the derivative operators act on $\alpha$  instead of $f^{AB}$. 

We begin with an examination of the first term in $\delta {\sf j}^{ab}$. After shifting the derivative operator away from $\Gamma^{AB}$, inserting Eq.~(\ref{bondiFE3b}), moving derivatives from $f^{AB}$ to $\alpha$, and making extensive use of Eqs.~(\ref{Oidentities}), we find that 
\begin{align} 
\Omega^{[a} \Omega^{b]}_A\, \alpha D_B (6 \Gamma^{AB}) &\sim 
\Omega^{[a} \Omega^{b]}_A \biggl( -4M \alpha^A -4 f^{AB} \alpha_B 
+ \frac{3}{8} \frac{\partial f^2}{\partial u} \alpha^A +  f^{AB} D_B D^2\alpha - f^{BC} D^A D_B \alpha_C\biggr) 
\nonumber \\ & \quad \mbox{} 
-4 \Omega^{[a}_A \Omega^{b]}_B\, f^{AC} D^B \alpha_C.
\end{align} 
In a similar way we obtain the following intermediate results:
\begin{subequations} 
	\begin{align} 
	\Omega^{[a} \Omega^{b]}_A\, D^A h  &\sim 0, \\ 
	6 \Omega^{[a} \Omega^{b]}_A  \bigl( D^A \alpha_B \bigr) \bigl( D_C f^{BC} \bigr) &\sim 
	6 \Omega^{[a} \Omega^{b]}_A\, f^{AB} \alpha_B - 6 \Omega^{[a}_A \Omega^{b]}_B f^{AC} D^B \alpha_C 
	- 6 \Omega^{[a} \Omega^{b]}_A\, f^{BC} D^A D_B \alpha_C, \\ 
	-3 \Omega^{[a} \Omega^{b]}_A\, \alpha^B D_B D_C f^{AC} &\sim 
3 \Omega^{[a} \Omega^{b]}_A\, f^{AB} \alpha_B 
+ 3 \Omega^{[a}_A \Omega^{b]}_B\, f^{AC} D^B \alpha_C 
- 3 \Omega^{[a} \Omega^{b]}_A\, f^{AB} D_B D^2\alpha,  \\ 
	-3 \Omega^{[a} \Omega^{b]}_A\, \bigl( D^2\alpha \bigr) \bigl( D_B f^{AB} \bigr) &\sim 
	3 \Omega^{[a} \Omega^{b]}_A\, f^{AB} D_B D^2\alpha, 
	\end{align} 
\end{subequations} 
where $h=h(u,\theta^A)$ is any scalar function. Collecting results and simplifying, we arrive at 
\begin{equation} 
-16\pi \delta {\sf j}^{ab} \sim \Omega^{[a} \Omega^{b]}_A \bigl(-\alpha s^A + 8M \alpha^A + 8f^{AB} \alpha_B 4 f^{AB} D_B D^2\alpha - 4 f^{BC} D^A D_B \alpha_C  \bigr) 
- 4 \Omega^{[a}_A \Omega^{b]}_B\, f^{AC} D^B \alpha_C , 
\label{delta_j} 
\end{equation}  
where $s^A$ was introduced in Eq.~(\ref{sA_def}). Integration over the unit two-sphere yields $\delta {\sf J}^{ab}$. 

\subsection{Balance law}

We differentiate Eq.~(\ref{delta_j}) with respect to $u$, and make use of Eq.~(\ref{bondiFE2}) to express $\dot{M}$ in terms of $\dot{f}^{AB}$. We simplify the result with 
\begin{equation} 
2 \Omega^{[a} \Omega^{b]}_A\, \alpha^A D_B D_C \dot{f}^{BC} \sim 
4 \Omega^{[a}_A \Omega^{b]}_B\, \dot{f}^{AC} D_C \alpha^B 
+ 2 \Omega^{[a} \Omega^{b]}_A\, \dot{f}^{BC} D^A D_B \alpha_C, 
\end{equation} 
and obtain 
\begin{equation} 
-16\pi \delta \frac{\partial {\sf j}^{ab}}{\partial u} 
\sim \Omega^{[a} \Omega^{b]}_A \bigl( -\alpha \dot{s}^A - \dot{f}_{BC} \dot{f}^{BC} \alpha^A 
+ 8 \dot{f}^{AB} \alpha_B \bigr) 
+ \Omega^{[a} \Omega^{b]}_A \bigl( 4 \dot{f}^{AB} D_B D^2\alpha - 2 \dot{f}^{BC} D^A D_B \alpha_C \bigr).   
\label{delta_jdot} 
\end{equation}  
Integration over the unit two-sphere gives $\delta (d {\sf J}^{ab}/du)$. 

Inspection of Eqs.~(\ref{delta_t}) and (\ref{delta_jdot}) reveals that both sides of the balance equation, 
\begin{equation} 
\frac{d}{du} {\sf J}^{ab} = -{\sf T}^{ab}, 
\end{equation} 
change under an infinitesimal supertranslation. The changes, however, are mutually consistent: we have that  
$\delta (d {\sf J}^{ab}/du) = -\delta {\sf T}^{ab}$ holds as a matter of identity, so that the balance law remains valid after the transformation. There is of course no surprise in this statement, because the balance equation is itself an identity derived from the Landau-Lifshitz formulation of the Einstein field equations.

\section{Tensor components} 
\label{app:comp} 

In this appendix we give an explicit listing of components for the tensors introduced in Sec.~\ref{sec:flux}. We recall that the angular-momentum flux integrand $\gotht^{ab}$ can be decomposed as $\gotht^{ab} = \partial \gothr^{ab}/\partial u + \goths^{ab}$. For the components of $\gothr^{a b}$ we have 
\begin{subequations} 
	\begin{align} 
	\gothr^{xy} &= \frac{1}{32\pi} \biggl[ \sin\theta\, Y \frac{\partial X}{\partial\theta} 
	+ X \frac{\partial X}{\partial \phi} - \sin\theta\, X \frac{\partial Y}{\partial\theta} 
	+ Y \frac{\partial Y}{\partial \phi} \biggr], \\ 
	\gothr^{yz} &= \frac{1}{32\pi\sin\theta} \biggl[ 
	-\sin\theta \bigl( \sin\phi\, X + \cos\theta\cos\phi\, Y \bigr) \frac{\partial X}{\partial\theta}  
	- \bigl( \cos\theta\cos\phi\, X - \sin\phi\, Y \bigr) \frac{\partial X}{\partial\phi} 
	\nonumber \\ & \quad \mbox{} 
	+ \sin\theta \bigl( \cos\theta\cos\phi\, X - \sin\phi\, Y \bigr) \frac{\partial Y}{\partial\theta}  
	- \bigl( \sin\phi\, X + \cos\theta \cos\phi\, Y \bigr) \frac{\partial Y}{\partial\phi} 
	- 2\cos\theta\sin\phi \bigl( X^2 + Y^2 \bigr) \biggr], \\ 
	\gothr^{zx} &= \frac{1}{32\pi\sin\theta} \biggl[ 
	\sin\theta \bigl( \cos\phi\, X - \cos\theta\sin\phi\, Y \bigr) \frac{\partial X}{\partial\theta}  
	- \bigl( \cos\theta\sin\phi\, X + \cos\phi\, Y \bigr) \frac{\partial X}{\partial\phi} 
	\nonumber \\ & \quad \mbox{} 
	+ \sin\theta \bigl( \cos\theta\sin\phi\, X + \cos\phi\, Y \bigr) \frac{\partial Y}{\partial\theta}  
	+ \bigl( \cos\phi\, X - \cos\theta \sin\phi\, Y \bigr) \frac{\partial Y}{\partial\phi} 
	+ 2\cos\theta\cos\phi \bigl( X^2 + Y^2 \bigr) \biggr]. 
	\end{align}
\end{subequations} 
For those of $\goths^{a b}$ we have 
\begin{subequations} 
	\begin{align} 
	\goths^{xy} &= \frac{1}{16\pi} \biggl\{
	-\biggl[ \frac{\partial X}{\partial\phi} - 2\sin\theta \frac{\partial Y}{\partial\theta} 
	- 2\cos\theta\, Y \biggr] \frac{\partial X}{\partial u} 
	- \biggl[ \frac{\partial Y}{\partial\phi} + 2\sin\theta \frac{\partial X}{\partial\theta} 
	+ 2\cos\theta\, X \biggr] \frac{\partial Y}{\partial u} \biggr\}, \\ 
	\goths^{yz} &= \frac{1}{16\pi\sin\theta} \biggl\{
	\biggl[ \sin\theta\sin\phi\, \frac{\partial X}{\partial\theta} 
	+ \cos\theta\cos\phi\, \frac{\partial X}{\partial\phi} 
	- 2\sin\theta\cos\theta\cos\phi\, \frac{\partial Y}{\partial\theta} 
	+ 2\sin\phi\, \frac{\partial Y}{\partial\phi} 
	\nonumber \\ & \quad \mbox{} 
	+ 2\cos\theta \bigl( 2\sin\phi\, X - \cos\theta\cos\phi\, Y \bigr) \biggr] \frac{\partial X}{\partial u} 
	\nonumber \\ & \quad \mbox{} 
	+ \biggl[ \sin\theta\sin\phi\, \frac{\partial Y}{\partial\theta} 
	+ \cos\theta\cos\phi\, \frac{\partial Y}{\partial\phi} 
	+ 2\sin\theta\cos\theta\cos\phi\, \frac{\partial X}{\partial\theta} 
	- 2\sin\phi\, \frac{\partial X}{\partial\phi} 
	\nonumber \\ & \quad \mbox{} 
	+ 2\cos\theta \bigl( 2\sin\phi\, Y + \cos\theta\cos\phi\, X \bigr) \biggr] \frac{\partial Y}{\partial u} 
	\biggr\}, \\ 
	\goths^{zx} &= \frac{1}{16\pi\sin\theta} \biggl\{
	\biggl[ -\sin\theta\cos\phi\, \frac{\partial X}{\partial\theta} 
	+ \cos\theta\sin\phi\, \frac{\partial X}{\partial\phi} 
	- 2\sin\theta\cos\theta\sin\phi\, \frac{\partial Y}{\partial\theta} 
	- 2\cos\phi\, \frac{\partial Y}{\partial\phi} 
	\nonumber \\ & \quad \mbox{} 
	- 2\cos\theta \bigl( 2\cos\phi\, X + \cos\theta\sin\phi\, Y \bigr) \biggr] \frac{\partial X}{\partial u} 
	\nonumber \\ & \quad \mbox{} 
	+ \biggl[ -\sin\theta\cos\phi\, \frac{\partial Y}{\partial\theta} 
	+ \cos\theta\sin\phi\, \frac{\partial Y}{\partial\phi} 
	+ 2\sin\theta\cos\theta\sin\phi\, \frac{\partial X}{\partial\theta} 
	+ 2\cos\phi\, \frac{\partial X}{\partial\phi} 
	\nonumber \\ & \quad \mbox{} 
	- 2\cos\theta \bigl( 2\cos\phi\, Y - \cos\theta\sin\phi\, X \bigr) \biggr] \frac{\partial Y}{\partial u} 
	\biggr\}. 
	\end{align} 
\end{subequations} 
The standard expression $\gotht^{a b}_{\rm standard}$ for the angular-momentum flux integrand comes with components 
\begin{subequations} 
	\begin{align} 
	\gotht^{xy}_{\rm standard} &= -\frac{1}{16\pi} \biggl[ 
	\frac{\partial X}{\partial\phi}\, \frac{\partial X}{\partial u} + 
	\frac{\partial Y}{\partial\phi}\, \frac{\partial Y}{\partial u} \biggr], \\ 
	\gotht^{yz}_{\rm standard} &= \frac{1}{16\pi\sin\theta} \biggl[ 
	\biggl(  \sin\theta\sin\phi\, \frac{\partial X}{\partial\theta} 
	+ \cos\theta\cos\phi\, \frac{\partial X}{\partial\phi} 
	- 2\cos\phi\, Y \biggr) \frac{\partial X}{\partial u} 
	\nonumber \\ & \quad \mbox{} 
	+ \biggl(  \sin\theta\sin\phi\, \frac{\partial Y}{\partial\theta} 
	+ \cos\theta\cos\phi\, \frac{\partial Y}{\partial\phi} 
	+ 2\cos\phi\, X \biggr) \frac{\partial Y}{\partial u} \biggr], \\ 
	\gotht^{zx}_{\rm standard} &= -\frac{1}{16\pi\sin\theta} \biggl[ 
	\biggl(  \sin\theta\cos\phi\, \frac{\partial X}{\partial\theta} 
	- \cos\theta\sin\phi\, \frac{\partial X}{\partial\phi} 
	+ 2\sin\phi\, Y \biggr) \frac{\partial X}{\partial u} 
	\nonumber \\ & \quad \mbox{} 
	+ \biggl(  \sin\theta\cos\phi\, \frac{\partial Y}{\partial\theta} 
	- \cos\theta\sin\phi\, \frac{\partial Y}{\partial\phi} 
	- 2\sin\phi\, X \biggr) \frac{\partial Y}{\partial u} \biggr]. 
	\end{align} 
\end{subequations} 
Part of the difference between $\gotht^{a b}_{\rm standard}$ and $\gotht^{a b}$ is encoded in $\gothp^{ab}$, whose components are
\begin{subequations} 
	\begin{align} 
	\gothp^{xy} &= \frac{1}{32\pi} \biggl[ 
	-\sin\theta\, Y \frac{\partial X}{\partial \theta} 
	+ 3 X \frac{\partial X}{\partial \phi} 
	+ \sin\theta\, X \frac{\partial Y}{\partial \theta}  
	+ 3 Y \frac{\partial Y}{\partial \phi} \biggr], \\ 
	\gothp^{yz} &= -\frac{1}{32\pi\sin\theta} \biggl[ 
	\sin\theta \bigl( 3\sin\phi\, X - \cos\theta\cos\phi\, Y \bigr) \frac{\partial X}{\partial\theta} 
	+ \bigl( 3\cos\theta\cos\phi\, X + \sin\phi\, Y \bigr) \frac{\partial X}{\partial\phi} 
	\nonumber \\ & \quad \mbox{} 
	+ \sin\theta \bigl( 3\sin\phi\, Y + \cos\theta\cos\phi\, X \bigr) \frac{\partial Y}{\partial\theta} 
	+ \bigl( 3\cos\theta\cos\phi\, Y - \sin\phi\, X \bigr) \frac{\partial Y}{\partial\phi} 
	- 2\cos\theta\sin\phi \bigl( X^2 + Y^2 \bigr) \biggr], \\ 
	\gothp^{zx} &= \frac{1}{32\pi\sin\theta} \biggl[ 
	\sin\theta \bigl( 3\cos\phi\, X + \cos\theta\sin\phi\, Y \bigr) \frac{\partial X}{\partial\theta} 
	- \bigl( 3\cos\theta\sin\phi\, X - \cos\phi\, Y \bigr) \frac{\partial X}{\partial\phi} 
	\nonumber \\ & \quad \mbox{} 
	+ \sin\theta \bigl( 3\cos\phi\, Y - \cos\theta\sin\phi\, X \bigr) \frac{\partial Y}{\partial\theta} 
	- \bigl( 3\cos\theta\sin\phi\, Y + \cos\phi\, X \bigr) \frac{\partial Y}{\partial\phi} 
	- 2\cos\theta\cos\phi \bigl( X^2 + Y^2 \bigr) \biggr]. 
	\end{align}
\end{subequations}  

\section{Boosted Schwarzschild metric} 
\label{app:boosted} 

In this appendix we identify the ``Coulombic pieces'' of the gravitational field of a boosted black hole, and show how they are encoded in the Bondi mass aspect, previously obtained in \cite{BvBM}. Moreover, we show that the angular-momentum integrand is nonzero for a boosted black hole, but that the total angular momentum properly vanishes. 

\subsection{Black hole frame} 

The metric of a Schwarzschild black hole in its own rest frame, expressed in Eddington-Finkelstein coordinates $(\bar{u},\bar{r},\bar{\theta},\bar{\phi})$, is given by   
\begin{equation} 
ds^2 = -(1-2m/\bar{r})\, d\bar{u}^2 - 2 d\bar{u}\bar{r} + \bar{r}^2\, d\bar{\Omega}^2,
\end{equation} 
where $m$ is the black hole's mass and $d\bar{\Omega}^2 := d\bar{\theta}^2 + \sin^2\bar{\theta}\, d\bar{\phi}^2$. In these coordinates the metric admits a Kerr-Schild form 
\begin{equation} 
g_{\bar{\alpha}\bar{\beta}} = g^0_{\bar{\alpha}\bar{\beta}} 
+ \frac{2m}{\bar{r}} k_{\bar{\alpha}} k_{\bar{\beta}}, 
\end{equation} 
where 
\begin{equation} 
g^0_{\bar{\alpha}\bar{\beta}}\, dx^{\bar{\alpha}} dx^{\bar{\beta}} 
= -d\bar{u}^2 - 2 d\bar{u}d\bar{r} + \bar{r}^2\, d\bar{\Omega}^2
\end{equation} 
is the metric of flat spacetime, and 
\begin{equation} 
k_{\bar{\alpha}}\, dx^{\bar{\alpha}} = -d\bar{u} 
\end{equation} 
is a null vector field. (The vector is null in both metrics.) The Kerr-Schild form is particularly convenient to obtain the boosted version of the metric.

\subsection{Laboratory frame} 

To the system $(\bar{u},\bar{r},\bar{\theta},\bar{\phi})$ we associate Lorentzian coordinates $(\bar{t},\bar{x},\bar{y},\bar{z})$, with $\bar{t} = \bar{u} + \bar{r}$ and $\bar{x}^a = \bar{r} \Omega^a(\bar{\theta}^A)$. We consider an observer that is boosted with respect to the black-hole frame, and attach a ``laboratory frame'' to this observer. The laboratory-frame coordinates $(t,x,y,z)$ are related to those of the black-hole frame by the Lorentz transformation 
\begin{equation} 
\bar{t} = \gamma(t-v\, z), \qquad \bar{x} = x, \qquad \bar{y} = y, \qquad 
\bar{z} = \gamma(z - v\, t), 
\end{equation} 
where $v$ is the boost velocity, and $\gamma := (1-v^2)^{-1/2}$. For simplicity, and without loss of generality, we choose the  boost to be directed along the $z$-axis. In the laboratory frame we introduce the retarded time $u := t - r$ and the spherical coordinates $(r,\theta^A)$ with $x^a = r \Omega^a(\theta^A)$. 

The metric of the boosted black hole continues to take the Kerr-Schild form 
\begin{equation} 
g_{\alpha\beta} = g^0_{\alpha\beta} + \frac{2m}{\bar{r}}\, k_\alpha k_\beta, 
\label{eq:bh-boosted}
\end{equation} 
where $g^0_{\alpha\beta}$ is the metric of flat spacetime and $k_\alpha$ is a null vector field. 

\subsection{Retarded coordinates} 

In the retarded coordinates $(u,r,\theta,\phi)$, the Minkowski metric in Eq.~(\ref{eq:bh-boosted}) is given by 
\begin{equation} 
g^0_{\alpha\beta}\, dx^\alpha dx^\beta = -du^2 - 2 dudr + r^2\, d\Omega^2,  
\end{equation} 
and the distance to the black hole is $\bar{r} = \gamma s(u,r,\theta)$ with 
\begin{equation} 
s := \bigl[ (1-v\cos\theta)^2 r^2 + 2v(v-\cos\theta)ur + v^2 u^2 \bigr]^{1/2}. 
\end{equation} 
We also have that $\bar{u} = \gamma[u + (1 - \cos\theta)v - s]$, and the nonvanishing components of 
$k_\alpha = -\partial_\alpha \bar{u}$ are   
\begin{subequations} 
\label{k_list} 
\begin{align} 
k_u &= -\gamma \biggl\{ 1 - \frac{v}{s} \Bigl[ (v-\cos\theta) r + v u \Bigr] \bigg\}, \\ 
k_r &= -\gamma \biggl\{ 1 - v\cos\theta 
- \frac{1}{s} \Bigl[ (1-v\cos\theta)^2 r + v(v-\cos\theta) u \Bigr] \bigg\}, \\
k_\theta &= -\gamma v r \sin\theta \biggl\{ 1 
- \frac{1}{s} \Bigl[ (1-v\cos\theta) r + u \Bigr] \biggr\}. 
\end{align} 
\end{subequations} 
From all this it follows that the metric components at large $r$ (and fixed $u$) are given by   
\begin{subequations} 
\begin{align} 
g_{uu} &= -1 + \frac{2m}{\gamma^3(1-v\cos\theta)^3} \frac{1}{r} + O(r^{-2}), \\ 
g_{ur} &= -1 + O(r^{-3}), \\ 
g_{u\theta} &= -\frac{2mv\sin\theta}{\gamma^3 (1-v\cos\theta)^4} \frac{u}{r} + O(r^{-2}), \\ 
g_{rr} &= O(r^{-5}), \\ 
g_{r\theta} &= O(r^{-3}), \\ 
g_{\theta\theta} &= r^2 + O(r^{-1}), \\ 
g_{\phi\phi} &= r^2\sin^2\theta + O(r^{-1}). 
\end{align} 
\end{subequations}  
We note that the $(u,r,\theta,\phi)$ coordinates are not Bondi coordinates --- $u$ is not null --- and the metric is therefore not in the Bondi form. This failure, however, is measured by $g^{uu} = O(r^{-5})$ and $g^{u\theta} = O(r^{-5})$, and we see that the Bondi form is recovered to a sufficient degree of accuracy to reveal the identity of the leading-order metric functions. A comparison with Eqs.~(\ref{bondi_metric}) and (\ref{Bondi_expansions}) reveals that $f_{AB} = 0 = A^A$ (as expected, given the absence of gravitational waves in this spacetime), and that 
\begin{equation} 
M = \frac{m}{\gamma^3(1-v\cos\theta)^3}, \qquad 
B^\theta = -\frac{2mv\sin\theta}{\gamma^3(1-v\cos\theta)^4} u. 
\end{equation} 
These quantities are related by the Einstein field equations: Eqs.~(\ref{bondiFE3a}) and (\ref{bondiFE3b}) imply that $\partial_u B^A = \frac{2}{3} \Omega^{AB} \partial_B M$ when $f_{AB} = 0$. The behavior of $B^A$ is therefore determined by $M$, and all ``Coulombic pieces'' of the gravitational field are encoded in the Bondi mass aspect $M$. 

\subsection{Angular momentum of the boosted black hole spacetime} 

Given that the boosted black hole does not radiate, we have that $f^{AB}=0$ and the flux of angular momentum vanishes. The angular-momentum integrand of Eq.~(\ref{j_final}), however, is non-zero. Inserting $f^{AB} = 0$ and the above expression for $B^A$ into $\gothj^{ab}_{\rm reg}$, we obtain
\begin{equation} 
\gothj^{xy}_{\rm reg} = 0, \qquad 
\gothj^{yz}_{\rm reg} = -\frac{3}{8\pi} \frac{mv}{\gamma^3} \frac{\sin\theta\sin\phi}{(1-v\cos\theta)^4}\, u,\qquad 
\gothj^{zx}_{\rm reg} = \frac{3}{8\pi} \frac{mv}{\gamma^3} \frac{\sin\theta\cos\phi}{(1-v\cos\theta)^4}\, u. 
\end{equation} 
This integrand grows linearly with $u$, but as expected, the integrated angular momentum vanishes: $J^{ab}=0$. 

\bibliography{references-ang-mom-2}

\begin{thebibliography}{25}%
\makeatletter
\providecommand \@ifxundefined [1]{%
 \@ifx{#1\undefined}
}%
\providecommand \@ifnum [1]{%
 \ifnum #1\expandafter \@firstoftwo
 \else \expandafter \@secondoftwo
 \fi
}%
\providecommand \@ifx [1]{%
 \ifx #1\expandafter \@firstoftwo
 \else \expandafter \@secondoftwo
 \fi
}%
\providecommand \natexlab [1]{#1}%
\providecommand \enquote  [1]{``#1''}%
\providecommand \bibnamefont  [1]{#1}%
\providecommand \bibfnamefont [1]{#1}%
\providecommand \citenamefont [1]{#1}%
\providecommand \href@noop [0]{\@secondoftwo}%
\providecommand \href [0]{\begingroup \@sanitize@url \@href}%
\providecommand \@href[1]{\@@startlink{#1}\@@href}%
\providecommand \@@href[1]{\endgroup#1\@@endlink}%
\providecommand \@sanitize@url [0]{\catcode `\\12\catcode `\$12\catcode
  `\&12\catcode `\#12\catcode `\^12\catcode `\_12\catcode `\%12\relax}%
\providecommand \@@startlink[1]{}%
\providecommand \@@endlink[0]{}%
\providecommand \url  [0]{\begingroup\@sanitize@url \@url }%
\providecommand \@url [1]{\endgroup\@href {#1}{\urlprefix }}%
\providecommand \urlprefix  [0]{URL }%
\providecommand \Eprint [0]{\href }%
\providecommand \doibase [0]{http://dx.doi.org/}%
\providecommand \selectlanguage [0]{\@gobble}%
\providecommand \bibinfo  [0]{\@secondoftwo}%
\providecommand \bibfield  [0]{\@secondoftwo}%
\providecommand \translation [1]{[#1]}%
\providecommand \BibitemOpen [0]{}%
\providecommand \bibitemStop [0]{}%
\providecommand \bibitemNoStop [0]{.\EOS\space}%
\providecommand \EOS [0]{\spacefactor3000\relax}%
\providecommand \BibitemShut  [1]{\csname bibitem#1\endcsname}%
\let\auto@bib@innerbib\@empty
\bibitem [{\citenamefont {Ashtekar}\ and\ \citenamefont
  {Bonga}(2017{\natexlab{a}})}]{ab1}%
  \BibitemOpen
  \bibfield  {author} {\bibinfo {author} {\bibfnamefont {A.}~\bibnamefont
  {Ashtekar}}\ and\ \bibinfo {author} {\bibfnamefont {B.}~\bibnamefont
  {Bonga}},\ }\href {\doibase 10.1007/s10714-017-2290-z} {\bibfield  {journal}
  {\bibinfo  {journal} {Gen. Rel. Grav.}\ }\textbf {\bibinfo {volume} {49}},\
  \bibinfo {pages} {122} (\bibinfo {year} {2017}{\natexlab{a}})},\ \Eprint
  {http://arxiv.org/abs/1707.09914} {arXiv:1707.09914} \BibitemShut {NoStop}%
\bibitem [{\citenamefont {Ashtekar}\ and\ \citenamefont
  {Bonga}(2017{\natexlab{b}})}]{ab2}%
  \BibitemOpen
  \bibfield  {author} {\bibinfo {author} {\bibfnamefont {A.}~\bibnamefont
  {Ashtekar}}\ and\ \bibinfo {author} {\bibfnamefont {B.}~\bibnamefont
  {Bonga}},\ }\href {\doibase 10.1088/1361-6382/aa88e2} {\bibfield  {journal}
  {\bibinfo  {journal} {Class. Quant. Grav.}\ }\textbf {\bibinfo {volume}
  {34}},\ \bibinfo {pages} {20LT01} (\bibinfo {year} {2017}{\natexlab{b}})},\
  \Eprint {http://arxiv.org/abs/1707.07729} {arXiv:1707.07729} \BibitemShut
  {NoStop}%
\bibitem [{\citenamefont {Bunster}\ \emph {et~al.}(2018)\citenamefont
  {Bunster}, \citenamefont {Gomberoff},\ and\ \citenamefont {P\'erez}}]{bgp}%
  \BibitemOpen
  \bibfield  {author} {\bibinfo {author} {\bibfnamefont {C.}~\bibnamefont
  {Bunster}}, \bibinfo {author} {\bibfnamefont {A.}~\bibnamefont {Gomberoff}},
  \ and\ \bibinfo {author} {\bibfnamefont {A.}~\bibnamefont {P\'erez}},\
  }\href@noop {} {\  (\bibinfo {year} {2018})},\ \Eprint
  {http://arxiv.org/abs/1805.03728} {arXiv:1805.03728 [hep-th]} \BibitemShut
  {NoStop}%
\bibitem [{\citenamefont {Bonga}\ \emph {et~al.}(2018)\citenamefont {Bonga},
  \citenamefont {Poisson},\ and\ \citenamefont {Yang}}]{bpy}%
  \BibitemOpen
  \bibfield  {author} {\bibinfo {author} {\bibfnamefont {B.}~\bibnamefont
  {Bonga}}, \bibinfo {author} {\bibfnamefont {E.}~\bibnamefont {Poisson}}, \
  and\ \bibinfo {author} {\bibfnamefont {H.}~\bibnamefont {Yang}},\ }\href@noop
  {} {\  (\bibinfo {year} {2018})},\ \Eprint {http://arxiv.org/abs/1805.01372}
  {arXiv:1805.01372} \BibitemShut {NoStop}%
\bibitem [{\citenamefont {Blanchet}(2014)}]{blanchet:14}%
  \BibitemOpen
  \bibfield  {author} {\bibinfo {author} {\bibfnamefont {L.}~\bibnamefont
  {Blanchet}},\ }\href@noop {} {\bibfield  {journal} {\bibinfo  {journal}
  {Living Reviews in Relativity}\ }\textbf {\bibinfo {volume} {17}},\ \bibinfo
  {pages} {2} (\bibinfo {year} {2014})}\BibitemShut {NoStop}%
\bibitem [{\citenamefont {Bondi}\ \emph {et~al.}(1962)\citenamefont {Bondi},
  \citenamefont {van~der Burg},\ and\ \citenamefont {Metzner}}]{BvBM}%
  \BibitemOpen
  \bibfield  {author} {\bibinfo {author} {\bibfnamefont {H.}~\bibnamefont
  {Bondi}}, \bibinfo {author} {\bibfnamefont {M.~G.~J.}\ \bibnamefont {van~der
  Burg}}, \ and\ \bibinfo {author} {\bibfnamefont {A.~W.~K.}\ \bibnamefont
  {Metzner}},\ }\href {\doibase 10.1098/rspa.1962.0223} {\bibfield  {journal}
  {\bibinfo  {journal} {Proc. Roy. Soc. Lond.}\ }\textbf {\bibinfo {volume}
  {A270}},\ \bibinfo {pages} {326} (\bibinfo {year} {1962})}\BibitemShut
  {NoStop}%
\bibitem [{\citenamefont {Thorne}(1980)}]{thorne}%
  \BibitemOpen
  \bibfield  {author} {\bibinfo {author} {\bibfnamefont {K.~S.}\ \bibnamefont
  {Thorne}},\ }\href {\doibase 10.1103/RevModPhys.52.299} {\bibfield  {journal}
  {\bibinfo  {journal} {Rev. Mod. Phys.}\ }\textbf {\bibinfo {volume} {52}},\
  \bibinfo {pages} {299} (\bibinfo {year} {1980})}\BibitemShut {NoStop}%
\bibitem [{\citenamefont {DeWitt}(2011)}]{dewitt}%
  \BibitemOpen
  \bibfield  {author} {\bibinfo {author} {\bibfnamefont {B.}~\bibnamefont
  {DeWitt}},\ }\href {\doibase 10.1007/978-3-540-36911-0} {\bibfield  {journal}
  {\bibinfo  {journal} {Lect. Notes Phys.}\ }\textbf {\bibinfo {volume}
  {826}},\ \bibinfo {pages} {pp.1} (\bibinfo {year} {2011})}\BibitemShut
  {NoStop}%
\bibitem [{\citenamefont {Poisson}\ and\ \citenamefont {Will}(2014)}]{pw}%
  \BibitemOpen
  \bibfield  {author} {\bibinfo {author} {\bibfnamefont {E.}~\bibnamefont
  {Poisson}}\ and\ \bibinfo {author} {\bibfnamefont {C.}~\bibnamefont {Will}},\
  }\href@noop {} {\emph {\bibinfo {title} {{Gravity, Newtonian, Post-Newtonian,
  Relativistic}}}}\ (\bibinfo  {publisher} {Cambridge University Press},\
  \bibinfo {year} {2014})\BibitemShut {NoStop}%
\bibitem [{\citenamefont {Ashtekar}\ and\ \citenamefont
  {Streubel}(1981)}]{AshtekarStreubel}%
  \BibitemOpen
  \bibfield  {author} {\bibinfo {author} {\bibfnamefont {A.}~\bibnamefont
  {Ashtekar}}\ and\ \bibinfo {author} {\bibfnamefont {M.}~\bibnamefont
  {Streubel}},\ }\href {\doibase 10.1098/rspa.1981.0109} {\bibfield  {journal}
  {\bibinfo  {journal} {Proc. Roy. Soc. Lond.}\ }\textbf {\bibinfo {volume}
  {A376}},\ \bibinfo {pages} {585} (\bibinfo {year} {1981})}\BibitemShut
  {NoStop}%
\bibitem [{\citenamefont {Wald}\ and\ \citenamefont
  {Zoupas}(2000)}]{WaldZoupas}%
  \BibitemOpen
  \bibfield  {author} {\bibinfo {author} {\bibfnamefont {R.~M.}\ \bibnamefont
  {Wald}}\ and\ \bibinfo {author} {\bibfnamefont {A.}~\bibnamefont {Zoupas}},\
  }\href {\doibase 10.1103/PhysRevD.61.084027} {\bibfield  {journal} {\bibinfo
  {journal} {Phys. Rev.}\ }\textbf {\bibinfo {volume} {D61}},\ \bibinfo {pages}
  {084027} (\bibinfo {year} {2000})},\ \Eprint
  {http://arxiv.org/abs/gr-qc/9911095} {arXiv:gr-qc/9911095 [gr-qc]}
  \BibitemShut {NoStop}%
\bibitem [{\citenamefont {Flanagan}\ and\ \citenamefont
  {Nichols}(2017)}]{FlanaganNichols}%
  \BibitemOpen
  \bibfield  {author} {\bibinfo {author} {\bibfnamefont {E.~E.}\ \bibnamefont
  {Flanagan}}\ and\ \bibinfo {author} {\bibfnamefont {D.~A.}\ \bibnamefont
  {Nichols}},\ }\href {\doibase 10.1103/PhysRevD.95.044002} {\bibfield
  {journal} {\bibinfo  {journal} {Phys. Rev.}\ }\textbf {\bibinfo {volume}
  {D95}},\ \bibinfo {pages} {044002} (\bibinfo {year} {2017})},\ \Eprint
  {http://arxiv.org/abs/1510.03386} {arXiv:1510.03386 [hep-th]} \BibitemShut
  {NoStop}%
\bibitem [{\citenamefont {Landau}\ and\ \citenamefont
  {Lifshitz}(2000)}]{landau-lifshitz:b2}%
  \BibitemOpen
  \bibfield  {author} {\bibinfo {author} {\bibfnamefont {L.~D.}\ \bibnamefont
  {Landau}}\ and\ \bibinfo {author} {\bibfnamefont {E.~M.}\ \bibnamefont
  {Lifshitz}},\ }\href@noop {} {\emph {\bibinfo {title} {The Classical Theory
  of Fields, Fourth Edition}}}\ (\bibinfo  {publisher}
  {Butterworth-Heinemann},\ \bibinfo {address} {Oxford, England},\ \bibinfo
  {year} {2000})\BibitemShut {NoStop}%
\bibitem [{\citenamefont {Sachs}(1962{\natexlab{a}})}]{Sachs}%
  \BibitemOpen
  \bibfield  {author} {\bibinfo {author} {\bibfnamefont {R.}~\bibnamefont
  {Sachs}},\ }\href {\doibase 10.1098/rspa.1962.0206} {\bibfield  {journal}
  {\bibinfo  {journal} {Proc. R. Soc. London, Ser. A}\ }\textbf {\bibinfo
  {volume} {270}},\ \bibinfo {pages} {103} (\bibinfo {year}
  {1962}{\natexlab{a}})}\BibitemShut {NoStop}%
\bibitem [{\citenamefont {Madler}\ and\ \citenamefont
  {Winicour}(2016)}]{bondisachsreview}%
  \BibitemOpen
  \bibfield  {author} {\bibinfo {author} {\bibfnamefont {T.}~\bibnamefont
  {Madler}}\ and\ \bibinfo {author} {\bibfnamefont {J.}~\bibnamefont
  {Winicour}},\ }\href {\doibase 10.4249/scholarpedia.33528} {\bibfield
  {journal} {\bibinfo  {journal} {Scholarpedia}\ }\textbf {\bibinfo {volume}
  {11}},\ \bibinfo {pages} {33528} (\bibinfo {year} {2016})},\ \Eprint
  {http://arxiv.org/abs/1609.01731} {arXiv:1609.01731 [gr-qc]} \BibitemShut
  {NoStop}%
\bibitem [{\citenamefont {Tamburino}\ and\ \citenamefont
  {Winicour}(1966)}]{tamburinowinicour}%
  \BibitemOpen
  \bibfield  {author} {\bibinfo {author} {\bibfnamefont {L.~A.}\ \bibnamefont
  {Tamburino}}\ and\ \bibinfo {author} {\bibfnamefont {J.~H.}\ \bibnamefont
  {Winicour}},\ }\href {\doibase 10.1103/PhysRev.150.1039} {\bibfield
  {journal} {\bibinfo  {journal} {Phys. Rev.}\ }\textbf {\bibinfo {volume}
  {150}},\ \bibinfo {pages} {1039} (\bibinfo {year} {1966})}\BibitemShut
  {NoStop}%
\bibitem [{\citenamefont {Compère}\ \emph {et~al.}(2018)\citenamefont
  {Compère}, \citenamefont {Fiorucci},\ and\ \citenamefont
  {Ruzziconi}}]{Compere:2018ylh}%
  \BibitemOpen
  \bibfield  {author} {\bibinfo {author} {\bibfnamefont {G.}~\bibnamefont
  {Compère}}, \bibinfo {author} {\bibfnamefont {A.}~\bibnamefont {Fiorucci}},
  \ and\ \bibinfo {author} {\bibfnamefont {R.}~\bibnamefont {Ruzziconi}},\
  }\href {\doibase 10.1007/JHEP11(2018)200} {\bibfield  {journal} {\bibinfo
  {journal} {JHEP}\ }\textbf {\bibinfo {volume} {11}},\ \bibinfo {pages} {200}
  (\bibinfo {year} {2018})},\ \bibinfo {note} {[JHEP18,200(2020)]},\ \Eprint
  {http://arxiv.org/abs/1810.00377} {arXiv:1810.00377 [hep-th]} \BibitemShut
  {NoStop}%
\bibitem [{grt()}]{grtensor}%
  \BibitemOpen
  \href@noop {} {}\bibinfo {note} {GrTensorIII, developed by Peter Musgrave,
  Denis Pollney, and Kayll Lake, is available free of charge at
  http://github.com/grtensor/grtensor}\BibitemShut {NoStop}%
\bibitem [{\citenamefont {Dray}\ and\ \citenamefont
  {Streubel}(1984)}]{DrayStreubel}%
  \BibitemOpen
  \bibfield  {author} {\bibinfo {author} {\bibfnamefont {T.}~\bibnamefont
  {Dray}}\ and\ \bibinfo {author} {\bibfnamefont {M.}~\bibnamefont
  {Streubel}},\ }\href {\doibase 10.1088/0264-9381/1/1/005} {\bibfield
  {journal} {\bibinfo  {journal} {Classical and Quantum Gravity}\ }\textbf
  {\bibinfo {volume} {1}},\ \bibinfo {pages} {15} (\bibinfo {year}
  {1984})}\BibitemShut {NoStop}%
\bibitem [{\citenamefont {Sachs}(1962{\natexlab{b}})}]{Sachs-BMS}%
  \BibitemOpen
  \bibfield  {author} {\bibinfo {author} {\bibfnamefont {R.}~\bibnamefont
  {Sachs}},\ }\href {\doibase 10.1103/PhysRev.128.2851} {\bibfield  {journal}
  {\bibinfo  {journal} {Phys. Rev.}\ }\textbf {\bibinfo {volume} {128}},\
  \bibinfo {pages} {2851} (\bibinfo {year} {1962}{\natexlab{b}})}\BibitemShut
  {NoStop}%
\bibitem [{\citenamefont {Nichols}(2018)}]{Nichols}%
  \BibitemOpen
  \bibfield  {author} {\bibinfo {author} {\bibfnamefont {D.~A.}\ \bibnamefont
  {Nichols}},\ }\href@noop {} {\  (\bibinfo {year} {2018})},\ \Eprint
  {http://arxiv.org/abs/1807.08767} {arXiv:1807.08767 [gr-qc]} \BibitemShut
  {NoStop}%
\bibitem [{\citenamefont {der Burg}\ and\ \citenamefont
  {Bondi}(1966)}]{BurgBondi}%
  \BibitemOpen
  \bibfield  {author} {\bibinfo {author} {\bibfnamefont {M.~G. J.~V.}\
  \bibnamefont {der Burg}}\ and\ \bibinfo {author} {\bibfnamefont
  {H.}~\bibnamefont {Bondi}},\ }\href {\doibase 10.1098/rspa.1966.0197}
  {\bibfield  {journal} {\bibinfo  {journal} {Proc. R. Soc. London, Ser. A}\
  }\textbf {\bibinfo {volume} {294}},\ \bibinfo {pages} {112} (\bibinfo {year}
  {1966})}\BibitemShut {NoStop}%
\bibitem [{\citenamefont {Newman}\ \emph {et~al.}(1968)\citenamefont {Newman},
  \citenamefont {Penrose},\ and\ \citenamefont {Bondi}}]{NewmanPenrose}%
  \BibitemOpen
  \bibfield  {author} {\bibinfo {author} {\bibfnamefont {E.~T.}\ \bibnamefont
  {Newman}}, \bibinfo {author} {\bibfnamefont {R.}~\bibnamefont {Penrose}}, \
  and\ \bibinfo {author} {\bibfnamefont {H.}~\bibnamefont {Bondi}},\ }\href
  {\doibase 10.1098/rspa.1968.0112} {\bibfield  {journal} {\bibinfo  {journal}
  {Proc. R. Soc. London, Ser. A}\ }\textbf {\bibinfo {volume} {305}},\ \bibinfo
  {pages} {175} (\bibinfo {year} {1968})}\BibitemShut {NoStop}%
\bibitem [{\citenamefont {Isaacson}\ and\ \citenamefont
  {Winicour}(1968)}]{isaacson-winicour:68}%
  \BibitemOpen
  \bibfield  {author} {\bibinfo {author} {\bibfnamefont {R.~A.}\ \bibnamefont
  {Isaacson}}\ and\ \bibinfo {author} {\bibfnamefont {J.}~\bibnamefont
  {Winicour}},\ }\href {\doibase 10.1103/PhysRev.168.1451} {\bibfield
  {journal} {\bibinfo  {journal} {Phys. Rev.}\ }\textbf {\bibinfo {volume}
  {168}},\ \bibinfo {pages} {1451} (\bibinfo {year} {1968})}\BibitemShut
  {NoStop}%
\bibitem [{\citenamefont {Blanchet}(1987)}]{blanchet:87}%
  \BibitemOpen
  \bibfield  {author} {\bibinfo {author} {\bibfnamefont {L.}~\bibnamefont
  {Blanchet}},\ }\href@noop {} {\bibfield  {journal} {\bibinfo  {journal}
  {Proc. R. Soc. London, Ser. A}\ }\textbf {\bibinfo {volume} {409}},\ \bibinfo
  {pages} {383–} (\bibinfo {year} {1987})}\BibitemShut {NoStop}%
\end{thebibliography}%

\end{document}